\documentclass[preprint,amsmath,showpacs]{revtex4}
\usepackage{graphicx}% Include figure files
\usepackage{dcolumn}% Align table columns on decimal point
\usepackage{bm}% bold math

\def\bee{\begin{eqnarray}}
\def\eee{\end{eqnarray}}

\begin{document}
\draft
\title{A semi-analytic calculation on the atmospheric tau neutrino flux in the GeV to TeV energy range}
\author{Fei-Fan Lee
and Guey-Lin Lin\footnote{Contact address: Institute of Physics,
National Chiao-Tung University, Hsinchu 300, Taiwan. E-mail:
glin@cc.nctu.edu.tw \ Tel: 886-3-5731984 \ Fax: 886-3-5720728} }
\address{
Institute of Physics, National Chiao-Tung University, Hsinchu 300,
Taiwan}
\date{\today}
\begin{abstract}
We present a semi-analytic calculation on the atmospheric tau
neutrino flux in the GeV to TeV energy range. The atmospheric
$\nu_{\tau}$ flux is calculated for the entire zenith angle range.
This flux is contributed by the oscillations of muon neutrinos
coming from the two-body $\pi, \ K$ decays and the three-body
$\mu^{\pm}$ decays, and the intrinsic tau neutrino flux surviving
the oscillations. The uncertainties in our calculations are
discussed in detail. The implications of our result are also
discussed.

\end{abstract}
\maketitle
\section{Introduction}
The flux of atmospheric tau neutrinos in the GeV to TeV energy range
comes from both the intrinsic atmospheric $\nu_{\tau}$ flux and the
flux due to the neutrino flavor oscillation $\nu_{\mu}\to
\nu_{\tau}$. The importance of understanding such a flux is twofold.
First, the detection of atmospheric $\nu_{\tau}$ flux is important
for confirming the atmospheric $\nu_{\mu}\to \nu_{\tau}$ oscillation
scenario which is so far established only by the $\nu_{\mu}$
disappearance measurement \cite{Ashie:2004mr}. Second, the
atmospheric $\nu_{\tau}$ flux is also an important background for
the search of astrophysical $\nu_{\tau}$ fluxes
\cite{Athar:2004pb,Athar:2004uk} or exotic $\nu_{\tau}$ fluxes such
as those arising from dark matter annihilations
\cite{Bertone:2004pz}, if an effective tau neutrino astronomy can be
developed in the future. The techniques for identifying the tau
neutrinos in the GeV to TeV energy range are discussed both in the
experiments for atmospheric tau neutrinos \cite{Stanev:1999ki} and
in the accelerator-based neutrino experiments
\cite{Zalewska:2004nd,Campagne:2004ay}. Due to growing attentions on
the direct $\nu_{\tau}$ detections, it is important to investigate
closely the flux of atmospheric tau neutrinos in such an energy
range.

The calculation of atmospheric $\nu_{\mu}$ and $\nu_e$ fluxes has
reached to a rather advanced stage \cite{Gaisser:2002jj}. The upward
atmospheric $\nu_{\tau}$ flux can be calculated easily from the
above $\nu_{\mu}$ flux by multiplying the $\nu_{\mu}\to \nu_{\tau}$
oscillation probability arising from neutrino propagation inside the
Earth. In such a case, the $\nu_{\mu}\to \nu_{\tau}$ oscillations in
the atmosphere is generally negligible. However, the atmospheric
neutrino coming with a zenith angle $\xi$ less than $90^{\circ}$
directly arrives at the detector. Hence oscillations of these
neutrinos in the atmosphere are precisely the problems one has to
deal with. In this regard, an estimation of the atmospheric
$\nu_{\tau}$ flux is given in Ref.~\cite{Athar:2004pb} for the
two-flavor neutrino oscillation framework \cite{Athar:2004mm}. A
detailed calculation of this flux in the same neutrino oscillation
framework is given in \cite{Athar:2004um} for zenith angles $0\leq
\xi \leq 60^{\circ}$, where the Earth curvature can be neglected in
the calculation. The extension of such a calculation to large zenith
angles is important. First of all, the path-length for the neutrino
propagation in the atmosphere increases drastically from
$\xi=60^{\circ}$ to $\xi=90^{\circ}$. As a result, the atmospheric
$\nu_{\tau}$ flux also increases drastically in this zenith angle
range by the enhanced $\nu_{\mu}\to \nu_{\tau}$ oscillation
probabilities. Secondly, the calculation of upward atmospheric
$\nu_{\tau}$ flux is also important since most of the neutrino
telescopes aim at detecting upward neutrino fluxes. In this paper,
we shall extend the calculation in \cite{Athar:2004um} to the entire
zenith angle range.

The calculations of atmospheric $\nu_{\tau}$ flux for zenith angles
$60^{\circ}\leq \xi\leq 90^{\circ}$ and zenith angles slightly
larger than $90^{\circ}$ are much more involved as we shall discuss
later. Besides extending our previous calculation to the entire
zenith angles, we also extend its validity from the energy range
$E_{\nu}\geq 10$ GeV to the energy range $E_{\nu}\geq 1$ GeV. This
improvement is accomplished by including the muon-decay contribution
$\mu^+\to \bar{\nu}_{\mu}\nu_e e^+$ and $\mu^-\to
\nu_{\mu}\bar{\nu}_e e^-$ to the intrinsic atmospheric $\nu_{\mu}$
flux, in additional to those arising from two-body $\pi$ and $K$
decays. Such a contribution also generates $\nu_{\tau}$ flux by
$\nu_{\mu}\to \nu_{\tau}$ oscillations. This part of $\nu_{\tau}$
flux is non-negligible for $E_{\nu}\leq 10$ GeV. Furthermore, it
contributes to the total $\nu_{\tau}$ flux in a growing percentage
as the zenith angle $\xi$ increases.

The paper is organized as follows. In Sec. II, we introduce the
method for calculating the intrinsic muon and tau neutrino fluxes.
Particularly we outline the strategy for dealing with atmospheric
neutrino flux for $\xi > 60^{\circ}$. In Sec. III, we present the
atmospheric tau neutrino flux taking into account the neutrino
flavor oscillations. We discuss implications of our results in Sec.
IV.

\section{The Intrinsic Atmospheric Neutrino Fluxes}
\subsection{Intrinsic atmospheric muon neutrino flux}
\begin{figure}
\includegraphics[width=7.in]{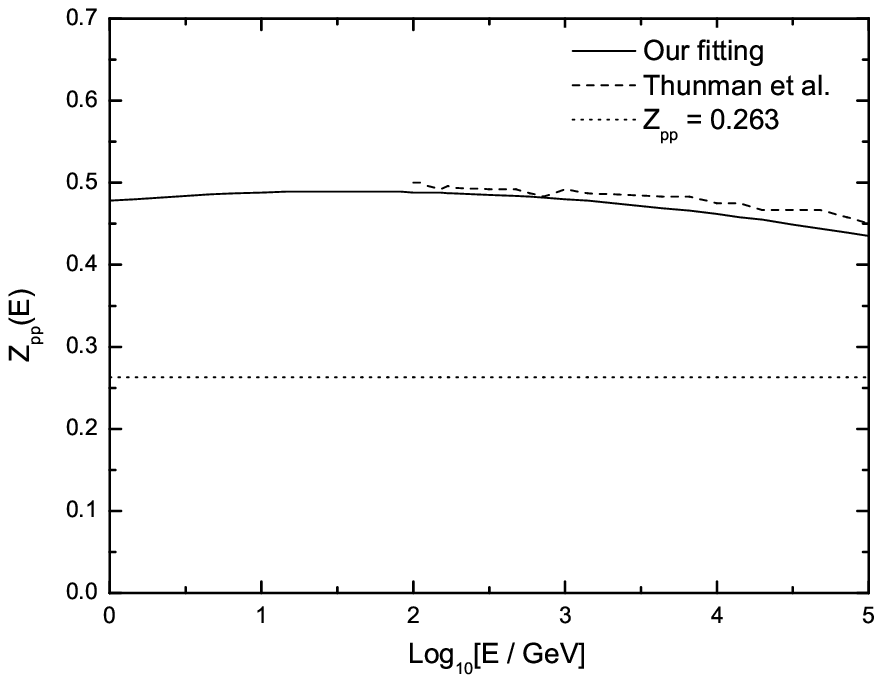}
\caption{The comparison of $Z_{pp}$ obtained by assuming the Feynman
scaling \cite{COSMIC} and that obtained by PYTHIA
\cite{Gondolo:1995fq}. Our extrapolation of the latter result is
also shown.} \label{zpp}
\end{figure}

We follow the approach in \cite{Gaisser:2001sd} for computing the
flux of intrinsic atmospheric muon neutrinos which could oscillate
into tau neutrinos. This approach computes the flux of muon
neutrinos coming from pion and kaon decays. The method for computing
muon neutrinos arising from muon decays will be discussed later. The
$\nu_{\mu}$ flux arising from $\pi$ decays reads:
\begin{eqnarray}
\label{atm-nu}
 \frac{\mbox{d}^2N^{\pi}_{\nu_{\mu}}(E,\xi,X)}{\mbox{d}E\mbox{d}X}&=&\int_E^{\infty}
 \mbox{d}E_N\int_E^{E_N}
 \mbox{d}E_{\pi}\frac{\Theta(E_{\pi}-\frac{E}{1-r_{\pi}})}{d_{\pi}E_{\pi}(1-r_{\pi})}
 \int_0^X
 \frac{\mbox{d}X'}{\lambda_N}P_{\pi}(E_{\pi},X,X')\nonumber \\
 & &\times \frac{1}{E_{\pi}}F_{N\pi}(E_{\pi},E_N)
 \times \exp \left(-\frac{X'}{\Lambda_N}\right)\phi_N(E_N),
\end{eqnarray}
where $E$ is the neutrino energy, $\xi$ is the zenith angle in the
direction of incident cosmic-ray nucleons,
$r_{\pi}=m_{\mu}^2/m_{\pi}^2$, $d_{\pi}$ is the pion decay length in
units of g/cm$^2$, $\lambda_N$ is the nucleon interaction length
while $\Lambda_N$ is the corresponding nucleon attenuation length,
and $\phi_N(E_N)$ is the primary cosmic-ray spectrum. For the
simplicity in discussions, we only consider the proton component of
$\phi_N$, which is given by \cite{Gaisser:2002jj}
\begin{equation}
\label{cosmic}
 \phi_p(E_{p})=1.49\cdot \left(E_{p}+2.15\cdot
 \exp(-0.21\sqrt{E_{p}})\right)^{-2.74},
\end{equation}
in units of cm$^{-2}$s$^{-1}$sr$^{-1}$GeV$^{-1}$. Since our
concerned energy range for the primary cosmic ray flux is between
$10^1$ GeV and $10^5$ GeV per nucleon (corresponding to roughly a
neutrino energy range between $10^0$ GeV and $10^4$ GeV), the
contribution by the heavier nuclei on the neutrino flux is between
$25\%$ and $40\%$ according to Fig. 7 of Ref.~\cite{Gaisser:2002jj}.
Hence one expects the eventual atmospheric $\nu_{\tau}$ flux is
underestimated by $25\%$ to $40\%$ by considering only the proton
component of the primary cosmic ray flux. The function
$P_{\pi}(E_{\pi},X,X')$ is the probability that a charged pion
produced at the slant depth $X'$ (g/cm$^2$) survives to the depth
$X$ ($
> X'$), $F_{N\pi}(E_{\pi},E_N)$ is the normalized inclusive cross
section for $N+{\rm air}\to \pi^{\pm}+Y$ given by
\cite{Gaisser:2001sd}:
\begin{equation}
\label{npi}
 F_{N\pi}(E_{\pi},E_N)\equiv
\frac{E_{\pi}}{\sigma_N}\frac{{\mbox d}\sigma(E_{\pi},E_N)}{{\mbox
d}E_{\pi}}=c_+(1-x)^{p_+}+c_-(1-x)^{p_-},
\end{equation}
where $x=E_{\pi}/E_N$, $c_+=0.92$, $c_-=0.81$, $p_+=4.1$, and
$p_-=4.8$. We remark that $c_+(1-x)^{p_+}$ corresponds to the
$\pi^+$ production while $c_-(1-x)^{p_-}$ corresponds to the $\pi^-$
production. The kaon contribution to the atmospheric $\nu_{\mu}$
flux has the same form as Eq.~(\ref{atm-nu}) with an inclusion of
the branching ratio $B(K\to \mu\nu)=0.635$ and appropriate
replacements in kinematic factors and the normalized inclusive cross
section. In particular, $F_{NK}(E_K,E_N)$ can be parameterized as
Eq.~(\ref{npi}) with $c_+=0.037$, $c_-=0.045$, $p_+=0.87$, and
$p_-=3.5$. Finally the nucleon interaction length, $\lambda_N$, and
the nucleon attenuation length, $\Lambda_N$, are both model
dependent. A simplified approach based upon the Feynman scaling
render both $\lambda_N$ and $\Lambda_N$ energy independent and
$Z_{pp}\equiv 1-\lambda_p/\Lambda_p=0.263$
\cite{COSMIC,Lipari:1993hd}, whereas a PYTHIA \cite{pythia}
calculation give rise to an energy dependent $Z_{pp}$
\cite{Gondolo:1995fq}. Both results on $Z_{pp}$ are compared in
Fig.~\ref{zpp} where we have extrapolated the energy dependent
$Z_{pp}(E)$ in Ref.~\cite{Gondolo:1995fq} down to $E=1$ GeV. The
above two approaches for calculating the hadronic $Z$ moments also
give rise to different results for $Z_{\pi\pi}$, $Z_{KK}$,
$Z_{N\pi}$ and $Z_{NK}$, where the last two $Z$-moments are related
to the productions of pions and kaons by the nucleon-air collisions.
In this paper, we shall only study the $Z_{pp}$ dependence of the
atmospheric $\nu_{\mu}$ flux (and consequently the atmospheric
$\nu_{\tau}$ flux) since the dependencies of this flux on $Z_{N\pi}$
and $Z_{NK}$ have been studied in \cite{Agrawal:1995gk}.
Furthermore, compared to the $Z_{pp}$ case, the values of
$Z_{\pi\pi}$ and $Z_{KK}$ obtained by the Feynman scaling do not
differ significantly from those obtained by the PYTHIA calculations,
as seen from \cite{Gondolo:1995fq}.

To proceed for calculating
$\mbox{d}^2N^{\pi}_{\nu_{\mu}}(E,\xi,X)/\mbox{d}E\mbox{d}X$, we note
that $P_{\pi}(E_{\pi},X,X')$ is given by \cite{Lipari:1993hd}
\begin{equation}
\label{pi_survive_a}
P_{\pi}(E_{\pi},X,X')=\exp\left(-\frac{X-X'}{\Lambda_{\pi}}\right)\cdot
\exp\left(-\frac{m_{\pi}c}{\tau_{\pi}}\int_{X'}^X\frac{{\mbox
d}T}{\rho(T)}\right),
\end{equation}
where $\Lambda_{\pi}=160$ g/cm$^2$ is the pion attenuation constant,
$\tau_{\pi}$ is the pion lifetime at its rest frame, while $\rho(T)$
is the atmosphere mass density at the slant depth $T$. For $\xi\leq
60^{\circ}$, the curvature of the Earth can be neglected so that
$\rho(T)=T\cos\xi/h_0$ with $h_0=6.4$ km the scale height for an
exponential atmosphere. In this approximation, the above survival
probability can be written as \cite{Gaisser:2001sd}
\begin{equation}
\label{pi_survive_b}
P_{\pi}(E_{\pi},X,X')=\exp\left(-\frac{X-X'}{\Lambda_{\pi}}\right)\cdot
\left(\frac{X'}{X}\right)^{\epsilon_{\pi}/E_{\pi}\cos\xi},
\end{equation}
where $\epsilon_{\pi}=m_{\pi}c^2h_0/c\tau_{\pi}$ is the pion decay
constant. Depending on the zenith angle, we apply either
Eq.~(\ref{pi_survive_a}) or Eq.~(\ref{pi_survive_b}) to perform the
calculations. The kaon survival probability $P_{K}(E_{K},X,X')$ has
the same form as $P_{\pi}(E_{\pi},X,X')$ except replacing
$\Lambda_{\pi}$ with $\Lambda_K$ and $\epsilon_{\pi}$ with
$\epsilon_K$. The two-body decay contribution to the atmospheric
$\nu_{\mu}$ flux is given by the sum of
$\mbox{d}^2N^{\pi}_{\nu_{\mu}}(E,\xi,X)/\mbox{d}E\mbox{d}X$ and
$\mbox{d}^2N^{K}_{\nu_{\mu}}(E,\xi,X)/\mbox{d}E\mbox{d}X$.

We recall that Eq.~(\ref{atm-nu}) and its corresponding form in the
kaon decay case only calculate the flux of muon neutrinos arising
from two-body pion and kaon decays. To calculate the contribution
from three-body muon decays, it is useful to first obtain the muon
flux \cite{Gaisser:2001sd}:
\begin{eqnarray}
\label{atm-mu}
\frac{\mbox{d}N^{\pi}_{\mu}(E,\xi,X)}{\mbox{d}E}&=&\int_{E'}^{\infty}
 \mbox{d}E_N\int_{E'}^{E_N}
 \mbox{d}E_{\pi}\int_0^{X}{\mbox
 d}X^{''}P_{\mu}(E,X,X^{''})\nonumber \\
 &\times& \frac{\Theta(E_{\pi}-E')\Theta(\frac{E'}{r_{\pi}}-E_{\pi})}{d_{\pi}E_{\pi}(1-r_{\pi})}
 \times \int_0^{X^{''}}
 \frac{\mbox{d}X'}{\lambda_N}
  P_{\pi}(E_{\pi},X^{''},X')\nonumber \\
 &\times& \frac{1}{E_{\pi}}F_{N\pi}(E_{\pi},E_N)
 \times \exp \left(-\frac{X'}{\Lambda_N}\right)\phi_N(E_N),
\end{eqnarray}
where $E'$ and $E$ are muon energies at slant depths $X^{''}$ and
$X$ respectively, while $P_{\mu}(E,X,X^{''})$ is the muon survival
probability given by \cite{Lipari:1993hd}
\begin{equation}
P_{\mu}(E,X,X^{''})=\exp\left[-\frac{m_{\mu}c}{\tau_{\mu}}\int_{X^{''}}^{X}{\mbox
d}T\frac{1}{E(T-X^{''},E')\rho(T)}\right],
\end{equation}
where $\tau_{\mu}$ is the muon lifetime at its rest frame and
$E(T-X^{''},E')$ is the muon energy at the slant depth $T$ with $E'$
the muon energy at its production point $X^{''}$.
%In the Earth's
%atmosphere, only the ionization process \cite{Rossi} is important
%for the muon energy loss so that
%\begin{equation}
%E(T-X^{''},E')=E'-\alpha(T-X^{''}),
%\end{equation}
%with $\alpha\approx 2$ MeV/g/cm$^2$ characterizing the muon
%ionization loss in the medium.
For the zenith angle $\xi\leq
60^{\circ}$, the above survival probability can be written as
\cite{Gaisser:2001sd}
\begin{equation}
P_{\mu}(E,X,X^{''})=\left(\frac{X^{''}}{X}\frac{E}{E+\alpha(X-X^{''})}\right)^
{\epsilon_{\mu}/(E\cos\xi+\alpha X\cos\xi)},
\end{equation}
with $\epsilon_{\mu}=m_{\mu}c^2h_0/c\tau_{\mu}$ the muon decay
constant and $\alpha\approx 2$ MeV/g/cm$^2$ characterizing the muon
ionization loss in the medium \cite{Rossi}. Since the muons are
polarized, it is convenient to keep track of the right-handed and
left-handed muon fluxes separately. The probability for a produced
$\mu^-$ to be right-handed or left-handed is determined by the muon
polarization \cite{Barr:1988rb,Barr:1989ru}:
\begin{equation}
\label{pol}
P_{\mu}(x)=\frac{1+r_{\pi}}{1-r_{\pi}}-\frac{2r_{\pi}}{(1-r_{\pi})x},
\end{equation}
with $x=E_{\mu}/E_{\pi}$ and $r_{\pi}=m_{\mu}^2/m_{\pi}^2$. Hence
$P_{R,L}(x)=\frac{1}{2}\left(1\pm P_{\mu}(x)\right)$ are the
probabilities for the produced muon to be right-handed or
left-handed respectively. The polarization for $\mu^+$ has an
opposite sign to that of $\mu^-$. The probabilities $P_{R,L}(x)$
should be inserted into Eq.~(\ref{atm-mu}) for obtaining four
different components of the muon flux:
$\mbox{d}N^{\pi}_{\mu^{+}_R}/\mbox{d}E$,
$\mbox{d}N^{\pi}_{\mu^{-}_R}/\mbox{d}E$,
$\mbox{d}N^{\pi}_{\mu^{+}_L}/\mbox{d}E$, and
$\mbox{d}N^{\pi}_{\mu^{-}_{L}}/\mbox{d}E$. There are additional four
components of the muon flux arising from the kaon decays. The
calculation of these components proceeds in the same way as the pion
decay case. The $\nu_{\mu}$ flux resulting from the muon flux is
then given by \cite{Lipari:1993hd}
\begin{equation}
\label{nu_mue}
\frac{\mbox{d}^2N^{\mu^{\pm}}_{\nu_{\mu}}(E,\xi,X)}{\mbox{d}E\mbox{d}X}=
\sum_{s=L,R}\int_E^{\infty}{\mbox d}E_{\mu}\frac{F_{\mu^{\pm}_s\to
\nu_{\mu}}(E/E_{\mu})}{d_{\mu}(E_{\mu},X)E_{\mu}}\cdot
\frac{\mbox{d}N_{\mu_s^{\pm}}(E_{\mu},\xi,X)}{\mbox{d}E_{\mu}},
\end{equation}
where $d_{\mu}(E_{\mu},X)$ is the muon decay length in units of
g/cm$^2$ at the slant depth $X$ and $F_{\mu^{\pm}_s\to
\nu_{\mu}}(E/E_{\mu})$ is the decay distribution of $\mu^{\pm}_s\to
\nu_{\mu}$. Precisely, in the ultra-relativistic limit, one has
\cite{Lipari:1993hd}
\begin{equation}
\label{distribution}
 F_{\mu^-\to \nu_{\mu}}(y)=g_0(y)+P_{\mu}g_1(y),
\end{equation}
with $g_0(y)=5/3-3y^2+4y^3/3$, $g_1(y)=1/3-3y^2+8y^3/3$. We do not
include the charm-hadron decay contribution to the muon neutrino
flux. It is shown in Ref.~\cite{Athar:2004um} that charm-hadron
decays contribute less than $5\%$ to the overall muon neutrino flux
for $E_{\nu}< 10^5$ GeV.

\subsection{Intrinsic atmospheric tau neutrino flux}
\begin{figure}
\includegraphics[width=7.in]{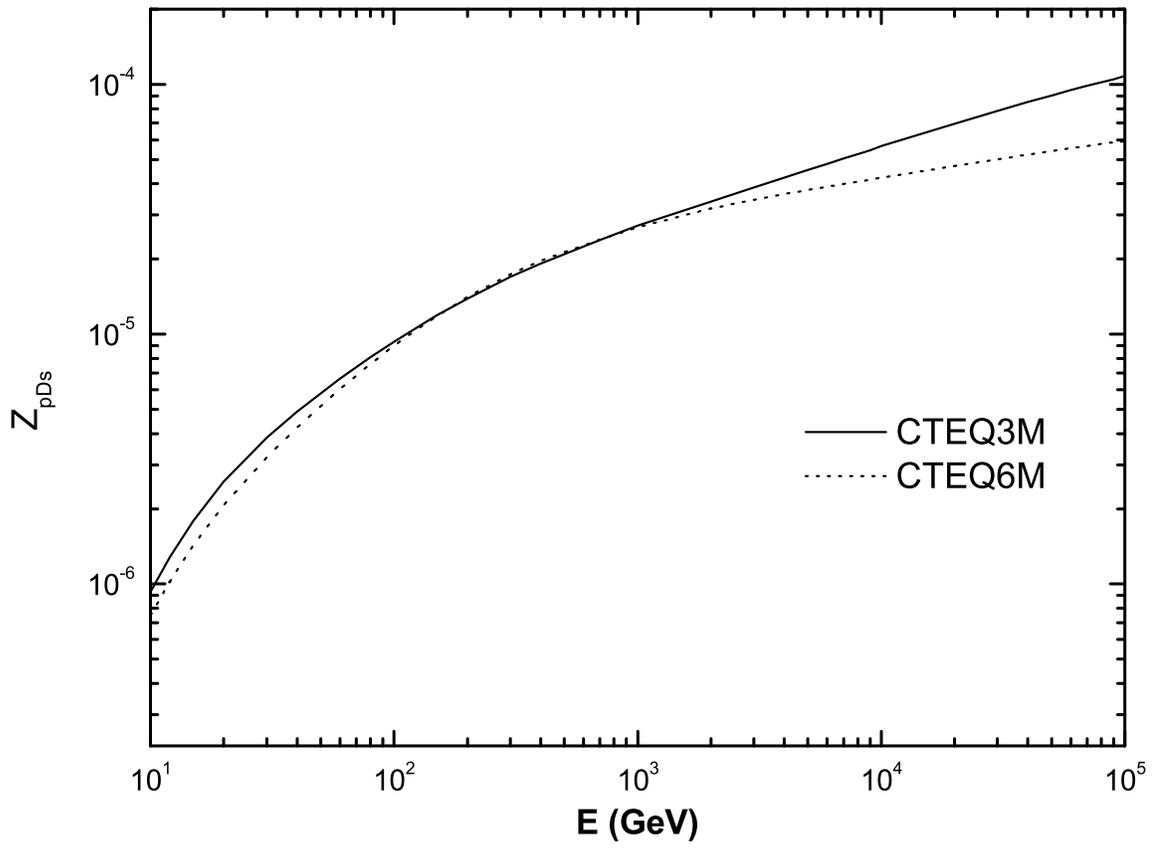}
\caption{The $Z$ moment $Z_{pD_s}$ obtained by perturbative QCD with
CTEQ3 and CTEQ6 parton distribution functions respectively.}
\label{zpds}
\end{figure}

To completely determine the atmospheric tau neutrino flux, we also
need to calculate its intrinsic component. Since the flux of
intrinsic atmospheric $\nu_{\tau}$ arises from $D_s$ decays, one
calculates this flux by solving the following cascade equations
\cite{COSMIC}:
\begin{eqnarray}
\frac{{\mbox d}\phi_p(E,X)}{{\mbox
d}X}&=&-\frac{\phi_p}{\lambda_p}+Z_{pp}\frac{\phi_p}{\lambda_p}\nonumber
\\
\frac{{\mbox d}\phi_{D_s}(E,X)}{{\mbox
d}X}&=&-\frac{\phi_{D_s}}{\lambda_{D_s}}-\frac{\phi_{D_s}}{d_{D_s}}+Z_{pD_s}\frac{\phi_p}{\lambda_p}
+Z_{D_sD_s}\frac{\phi_{D_s}}{\lambda_{D_s}}\nonumber \\
\frac{{\mbox d}\phi_{\nu_{\tau}}(E,X)}{{\mbox
d}X}&=&Z_{D_s\nu_{\tau}}\frac{\phi_{D_s}}{d_{D_s}}, \label{cascade}
\end{eqnarray}
 where the particle flux ${\mbox d}\phi_i(E,X)/{\mbox d}X$ denotes ${\mbox d}^2N_i(E,X)/{\mbox d}E{\mbox
 d}X$, $d_i$ and $\lambda_i$ denote particle's decay and interaction
 length in g/cm$^2$ respectively, and the $Z$ moments $Z_{ij}$ are
 defined by
\begin{equation}
\label{z-moment}
 Z_{ij}(E_j)\equiv \int_{E_j}^{\infty}{\mbox
 d}E_i\frac{\phi_i(E_i)}{\phi_i(E_j)}\frac{\lambda_i(E_j)}{\lambda_i(E_i)}
 \frac{{\mbox d}n_{iA\to jY}(E_i,E_j)}{{\mbox d}E_j},
\end{equation}
with ${\mbox d}n_{iA\to jY}(E_i,E_j)\equiv {\mbox d}\sigma_{iA\to
jY}(E_i,E_j)/\sigma_{iA}(E_i)$. In the decay process, the scattering
length $\lambda_i$ is replaced by the decay length $d_i$ while
${\mbox d}n_{iA\to jY}(E_i,E_j)/{\mbox d}E_j$ is replaced by the
decay distribution $F_{i\to j}(E_j/E_i)$. In our concerned energy
range, the chain of equations in (\ref{cascade}) can be easily
solved by simplifying the second equation, namely by neglecting
terms $\phi_{D_s}/\lambda_{D_s}$ and
$Z_{D_sD_s}\phi_{D_s}/\lambda_{D_s}$. One obtains
\begin{equation}
\label{atm-tau}
 \frac{\mbox{d}^2N_{\nu_{\tau}}(E,X)}{\mbox{d}E\mbox{d}X}=
 \frac{Z_{pD_s}(E)Z_{D_s\nu_{\tau}}(E)}{1-Z_{pp}(E)}\cdot
 \frac{\exp(-X/\Lambda_p)\phi_p(E)}{\Lambda_p}.
 \end{equation}
We use two different values for $Z_{pp}\equiv 1-\lambda_p/\Lambda_p$
as shown in Fig.~\ref{zpp}. To determine $Z_{pD_s}$, it is necessary
to calculate ${\mbox d}\sigma_{pA\to D_sY}(E_p,E_{D_s})/{\mbox
d}E_{D_s}$. Since $D_s$ meson is heavy enough, the above
differential cross section is calculable using perturbative QCD
\cite{Pasquali:1998xf}. In this work, the next-to-leading order
(NLO) perturbative QCD \cite{Nason:1989zy,Mangano:1991jk} with CTEQ6
parton distribution functions \cite{Pumplin:2002vw} are employed to
calculate the differential cross section of $pA\to c\bar{c}$. To
obtain ${\mbox d}\sigma_{pA\to D_sY}(E_p,E_{D_s})/{\mbox d}E_{D_s}$,
we multiply the charm quark differential cross section by the
probability factor $13\%$ to account for the fragmentation process
$c\to D_s$ \cite{Pasquali:1998xf}. In Fig.~\ref{zpds}, we compare
our $Z_{pD_s}$ to a previous result obtained by the CTEQ3 parton
distribution functions \cite{Pasquali:1998ji}. In the latter work,
the NLO pertubative QCD effects are taken into account by the $K$
factor defined by
\begin{eqnarray}
K(E,x_E)=\frac{\mbox{d}\sigma^{\rm
NLO}(E,x_E)/{\mbox{d}x_E}}{\mbox{d}\sigma^{\rm
LO}(E,x_E)/{\mbox{d}x_E}},
\end{eqnarray}
where $\mbox{d}\sigma^{\rm LO}(E,x_E)/{\mbox{d}x_E}$ and
$\mbox{d}\sigma^{\rm NLO}(E,x_E)/{\mbox{d}x_E}$ are leading order
and next-to-leading order differential cross sections for $pA\to
c\bar{c}$ respectively, with $x_E=E_c/E_p$. For QCD renormalization
scale $\mu=m_c$ and the factorization scale $M=2m_c$, the $K$ factor
is fitted to be \cite{Pasquali:1998ji}
\begin{eqnarray}
K(E,x_E)&=&1.36+0.42\ln\left(\ln(E/{\rm GeV})\right)\nonumber
\\
&+& \left(3.40+18.7(E/{\rm GeV})^{-0.43}-0.079\ln(E/{\rm
GeV})\right)x_E^{1.5}.
\end{eqnarray}
We apply this $K$ factor to our calculation with CTEQ6 parton
distribution functions. Comparing this result with that obtained by
applying CTEQ3 parton distribution functions, one acquires an idea
on the uncertainty of perturbative QCD approach to the charm hadron
production cross section. It is seen from Fig.~\ref{zpds} that both
$Z$ moments agree well for energies below TeV. For $E=10$ TeV, they
differ by about $30\%$.

Besides perturbative QCD approach, there are non-perturbative
approach for computing the charm hadron production cross section. In
fact, such non-perturbative approaches
\cite{Kaidalov:1985jg,Bugaev:1998bi} are motivated to accommodate
accelerator data on strange particle productions, which are
underestimated by the perturbative QCD approach. It is desirable to
apply these approaches to charm hadron productions. The
quark-gluon-string-model (QGSM) \cite{Kaidalov:1985jg} is a
non-perturbative approach based upon the string fragmentation, where
the model parameters are tuned to the production cross section of
strange particles. The recombination-quark-parton-model (RQPM)
\cite{Bugaev:1998bi} is also a phenomenological approach which takes
into account the contribution of the intrinsic charm in the nucleon
to the charm hadron production cross section. Detailed comparisons
of these two models with perturbative QCD approach on the charm
hadron productions are given in~\cite{Costa:2000jw}. It is shown
that perturbative QCD approach gives the smallest charm production
$Z$ moments. It is clear that the model dependencies on the charm
hadron productions affect both the prompt atmospheric muon neutrino
flux and the intrinsic atmospheric tau neutrino flux. A detailed
study on the model dependencies of the intrinsic atmospheric tau
neutrino flux is given in~\cite{Costa:2001fb}. We shall further
discuss these model dependencies after commenting on the $Z$ moment
$Z_{D_s\nu_{\tau}}$.

We note that $Z_{D_s\nu_{\tau}}$ is related to the energy
distributions of the $D_s$ decays into tau neutrinos. One arises
from the decay $D_s\to \nu_{\tau}\tau$, and the other follows from
the subsequent tau-lepton decay, $\tau\to \nu_{\tau}+X$. The latter
contribution is calculated using the decay distributions of the
decay modes $\tau\to \nu_{\tau}\rho$, $\tau\to \nu_{\tau}\pi$,
$\tau\to \nu_{\tau}a_1$ \cite{Li:1995aw,Pasquali:1998xf}, and
$\tau\to \nu_{\tau}l\nu_l$ \cite{COSMIC,Lipari:1993hd}.

\begin{figure}
\includegraphics[width=7.in]{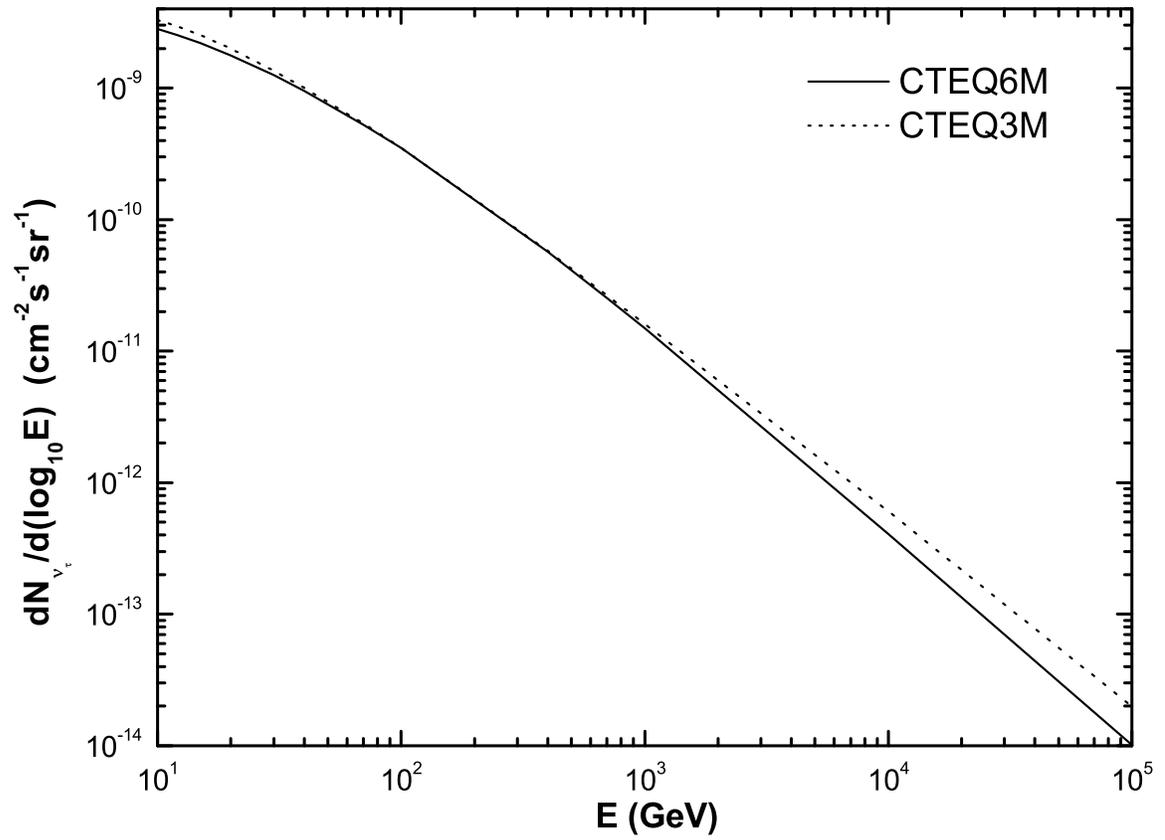}
\caption{The comparison of intrinsic atmospheric $\nu_{\tau}$ fluxes
calculated by perturbative QCD with CTEQ3 and CTEQ6 parton
distribution functions respectively. } \label{pdf}
\end{figure}

\begin{figure}
\includegraphics[width=7.in]{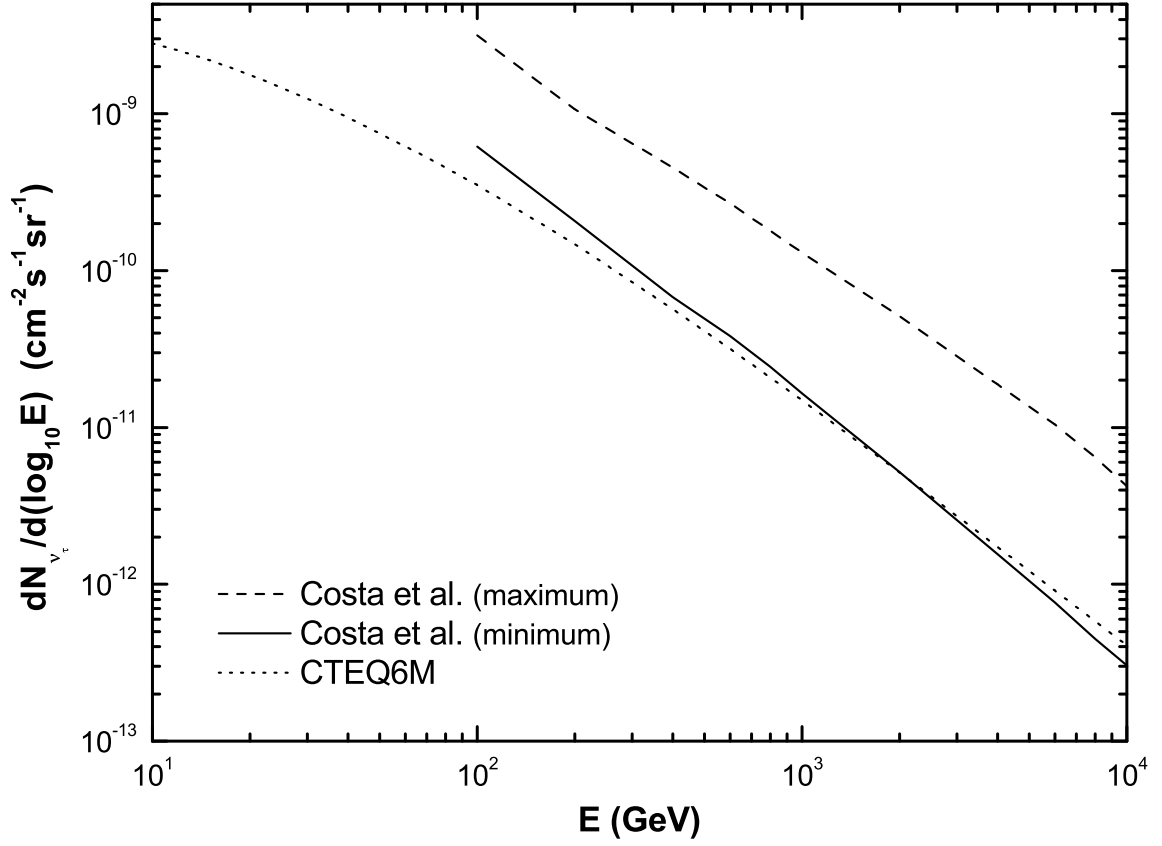}
\caption{The model dependencies of intrinsic atmospheric
$\nu_{\tau}$ flux. The minimal flux from Ref.~\cite{Costa:2001fb} is
given by perturbative QCD. The maximal flux from the same reference
is given by RQPM model for $E\leq 300$ GeV, and by QGSM model for
$E> 300$ GeV. } \label{tau_intrinsic}
\end{figure}

The uncertainty of intrinsic atmospheric $\nu_{\tau}$ flux due to
different approaches for $Z_{pp}$ is negligible. The main
uncertainty of this flux is due to the model dependence of the
$Z$-moment $Z_{pD_s}$. Within the perturbative QCD approach, the
dependence of this flux on the parton distribution functions is
shown in Fig.~\ref{pdf}. It is easily seen that the intrinsic
atmospheric $\nu_{\tau}$ flux is not sensitive to parton
distribution functions for $E< 10^3$ GeV. However, at $E=10^4$ GeV,
both fluxes differ by almost $50\%$. Incorporating the
non-perturbative approaches for charm hadron productions
\cite{Kaidalov:1985jg,Bugaev:1998bi}, the uncertainties of intrinsic
atmospheric $\nu_{\tau}$ flux is depicted in
Fig.~\ref{tau_intrinsic}. It is seen that the minimal $\nu_{\tau}$
flux in Ref.~\cite{Costa:2001fb} is consistent with our $\nu_{\tau}$
flux calculated by perturbative QCD with CTEQ6 parton distribution
functions. On the other hand, the maximal flux shown in
Fig.~\ref{tau_intrinsic} is almost one order of magnitude larger
than the minimal one. This maximal flux is given by the RQPM model
below 300 GeV while it is given by the QGSM model beyond this energy
\cite{Costa:2001va}. We remark that the original minimal and maximal
$\nu_{\tau}$ fluxes in Ref.~\cite{Costa:2001fb} correspond to
different sets of primary cosmic ray flux, which is considered as
one of the uncertainties for the $\nu_{\tau}$ flux. However, we have
re-scaled these fluxes to a common cosmic ray flux,
Eq.~(\ref{cosmic}), used in this paper. We also note that the
uncertainty of intrinsic atmospheric $\nu_{\tau}$ flux provided by
Ref.~\cite{Costa:2001fb} starts at $E = 100$ GeV, while our
calculation of this flux starts at $E=10$ GeV.

It is interesting to see how much the uncertainty of the intrinsic
$\nu_{\tau}$ flux could affect the determination of the $\nu_{\tau}$
flux taking into account the oscillation effect. In the next
section, we shall study this issue with respect to the upward
atmospheric $\nu_{\tau}$ flux where the oscillation effect is the
largest.
\section{The Atmospheric Tau Neutrino Flux with Oscillations}
\subsection{The Downward and Horizontal Atmospheric Tau Neutrino Fluxes}
The atmospheric tau neutrino flux can be calculated using
\begin{eqnarray}
\label{oscillate}
\frac{\mbox{d}\bar{N}_{\nu_{\tau}}(E,\xi)}{\mbox{d}E}&=&\int_0^{X_{\rm
max}(\xi)} {\mbox d}X\left[
\frac{\mbox{d}^2N_{\nu_{\mu}}(E,\xi,X)}{\mbox{d}E\mbox{d}X}\cdot
P_{\nu_{\mu}\to \nu_{\tau}}\left(E,L(X,\xi)\right)\right.\nonumber
\\
&&\left.+\frac{\mbox{d}^2N_{\nu_{\tau}}(E,\xi,X)}{\mbox{d}E\mbox{d}X}
\cdot \left(1-P_{\nu_{\mu}\to
\nu_{\tau}}\left(E,L(X,\xi)\right)\right)\right],
\end{eqnarray}
where $P_{\nu_{\mu}\to \nu_{\tau}}\left(E,L(X,\xi)\right)\equiv
\sin^22\theta_{23}\sin^2(1.27\Delta m_{31}^2L/E)$ is the
$\nu_{\mu}\to \nu_{\tau}$ oscillation probability, assuming a
vanishing $\theta_{13}$. We have used the notation
$\mbox{d}\bar{N}_{\nu_{\tau}}(E,\xi)/\mbox{d}E$ to denote the
atmospheric $\nu_{\tau}$ flux taking into account the oscillation
effect. The unit of $\Delta m_{31}^2$ is eV$^2$ while $L$ and $E$
are in units of km and GeV respectively. A recent SK analysis of the
atmospheric neutrino data implies \cite{Ashie:2004mr}
\begin{equation}
 \Delta m_{31}^{2}=(1.9-3.0)\cdot 10^{-3}\, \, \, {\rm eV}^{2}, \, \, \,
 \sin^{2}2\theta_{23} >0.9.
\label{range}
\end{equation}
This is a  $90\% \, {\rm C.L.}$ range with the best fit values given
by $\Delta m_{31}^{2}=2.4\cdot 10^{-3}\, \, \, {\rm eV}^{2}$ and
$\sin^{2}2\theta_{23} =1$ respectively.

\subsubsection{Meson decay contributions} 
\begin{figure}
\includegraphics[width=7.in]{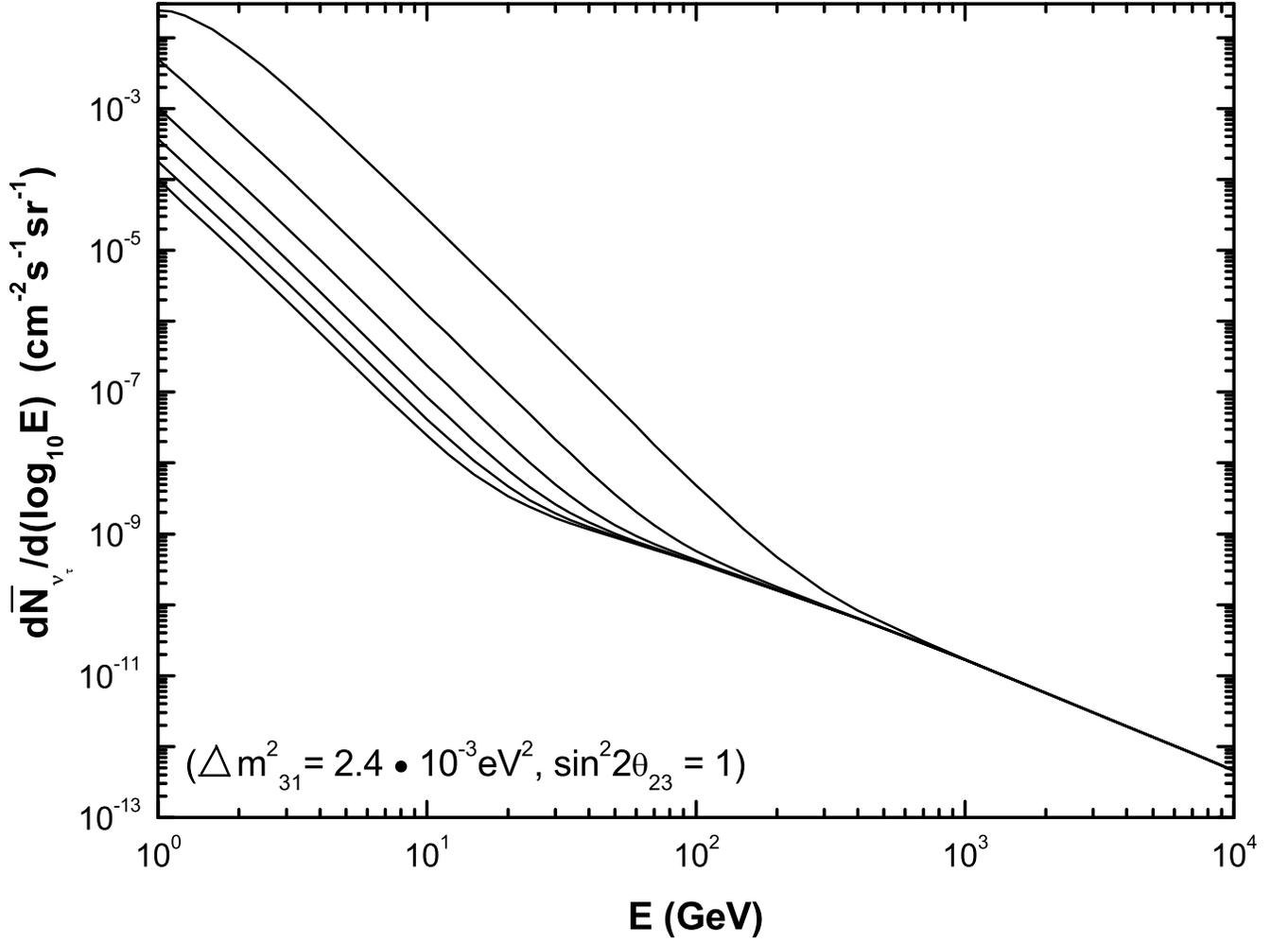}
\caption{The atmospheric $\nu_{\tau}$ flux for $\cos\xi=0, \ 0.2, \
0.4, \ 0.6, \ 0.8$ and $1$ (from top to bottom) with
$\sin^22\theta_{23}=1$, $\Delta m_{31}^{2}=2.4\cdot 10^{-3}\, \, \,
{\rm eV}^{2}$, and $Z_{pp}=0.263$.} \label{tau_flux}
\end{figure}

\begin{figure}
\includegraphics[width=7.in]{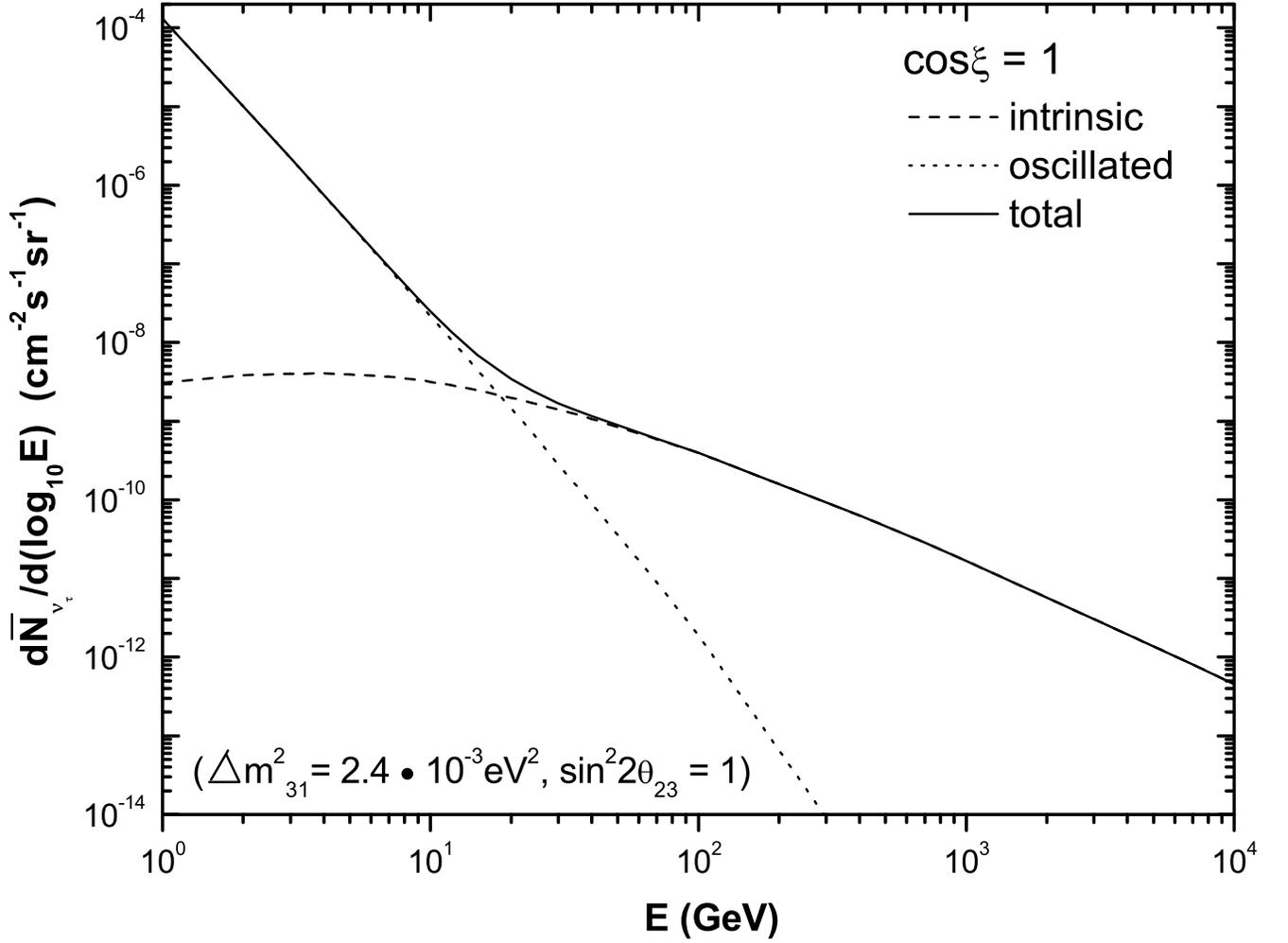}
\caption{The downward atmospheric $\nu_{\tau}$ flux (solid line) as
the sum of its oscillated (dotted line) and intrinsic (dashed line)
components. } \label{down_compare}
\end{figure}

Using the best fit values
of neutrino oscillation parameters, we obtain atmospheric tau
neutrino fluxes for $\cos\xi=0.2, \ 0.4, \ \cdots, \ 1$ as depicted
in Fig.~\ref{tau_flux}. This set of result is obtained by using an
energy-independent $Z$ moment, $Z_{pp}\equiv
1-\lambda_p/\Lambda_p=0.263$ mentioned earlier. For the $\nu_{\mu}$
flux on the R.H.S. of Eq.~(\ref{oscillate}), we only include the
two-body pion and kaon decay contributions. The muon-decay
contribution to this flux will be presented later. The intrinsic
$\nu_{\tau}$ flux in the same equation is taken to be that
calculated by perturbative QCD with CTEQ6 parton distribution
functions \cite{Pumplin:2002vw}. It is instructive to separately
present the oscillated and intrinsic atmospheric $\nu_{\tau}$ fluxes
corresponding to the two terms on the R.H.S. of
Eq.~(\ref{oscillate}). This is done for $\xi=0^{\circ}$ in
Fig.~\ref{down_compare}. We see that the oscillated and intrinsic
atmospheric $\nu_{\tau}$ fluxes cross at $E\simeq 20$ GeV,
indicating that the $\nu_{\mu}\to \nu_{\tau}$ oscillation effect
becomes important for $E< 20$ GeV for such a zenith angle.

We note that the atmospheric $\nu_{\tau}$ flux increases as $\xi$
increases from $0^{\circ}$ to $90^{\circ}$. There are two crucial
factors dictating the angular dependence of such a flux. First, the
atmosphere depth traversed by the cosmic ray particles increases as
the zenith angle $\xi$ increases. Second, the atmospheric muon
neutrinos are on-average produced more far away from the ground
detector for a larger zenith angle, implying a larger $\nu_{\mu}\to
\nu_{\tau}$ oscillation probability. In fact, the neutrino
path-length dependencies on the zenith angle $\xi$ and the neutrino
energy $E$ have been studied carefully by the Monte-Carlo simulation
\cite{Gaisser:1997eu}. Our semi-analytic approach reproduces these
dependencies very well. It is found that, for $E=10$ GeV and
$\cos\xi=1$ ($\xi=0^{\circ}$), the average neutrino path-length from
the $\nu_{\mu}$ production point to the ground detector is $14$ km.
The average neutrino path-length increases to $45$ km and $650$ km
for $\xi=66^{\circ}$ ($\cos\xi=0.4$) and $\xi=90^{\circ}$
respectively. The huge path-length of horizontal neutrinos makes the
$\nu_{\tau}$ flux in this direction two orders of magnitude larger
than the downward $\nu_{\tau}$ flux. It is also interesting to note
that the horizontal $\nu_{\tau}$ flux for $E$ approaching $1$ GeV
begins to show oscillatory behavior. This is because, for $E=1$ GeV
and $\Delta m_{31}^2=2.4\cdot 10^{-3}$ eV$^2$, $L_{\rm osc}\equiv
4E/\Delta m_{31}^2\approx 330$ km which is already shorter than the
average neutrino path-length at this zenith angle.

\begin{figure}
\includegraphics[width=7.in]{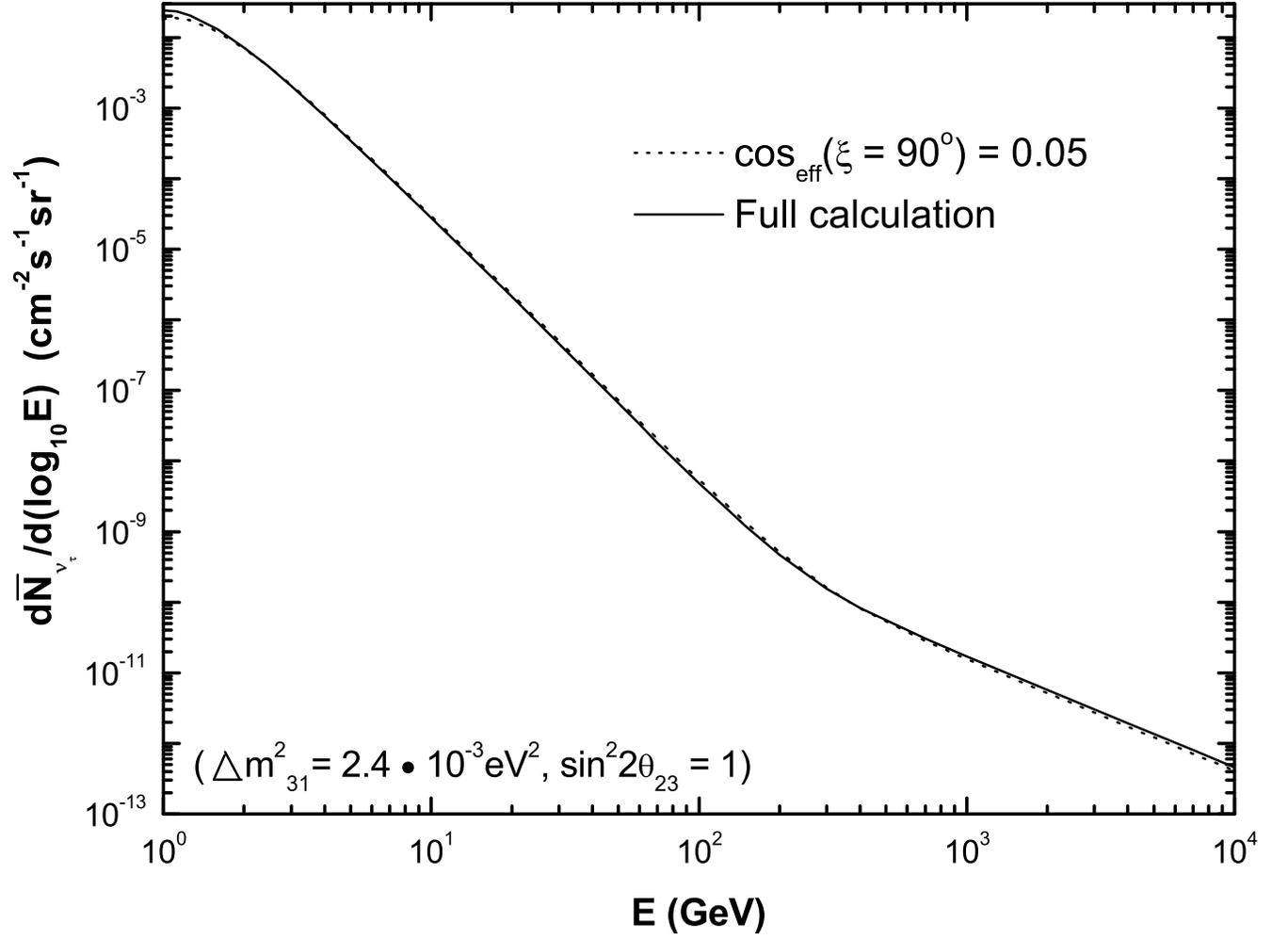}
\caption{The comparison of atmospheric $\nu_{\tau}$ flux obtained by
using $\cos_{\rm eff}\xi$ and that obtained by the full calculation
for $\xi=90^{\circ}$.} \label{comp}
\end{figure}

We stress that our calculation procedures for $\cos\xi > 0.5$ and
$\cos\xi < 0.5$ are different. In the former case, the curvature of
the Earth can be neglected and the pion or kaon survival probability
in the atmosphere is approximated by Eq.~(\ref{pi_survive_b}). This
is the approach we adopted in Ref.~\cite{Athar:2004um}. For
$\cos\xi< 0.5$, i.e., $\xi
> 60^{\circ}$, we use Eq.~(\ref{pi_survive_a}) for the meson survival
probability. In this case the calculation is much more involved as
the meson survival probability in Eq.~(\ref{pi_survive_a}) contains
an additional integration. It has been pointed out in
Ref.~\cite{Gaisser:1997eu} that one may apply
Eq.~(\ref{pi_survive_b}) for calculating the path-length
distribution of neutrinos for $\xi
> 60^{\circ}$ so long as one replaces $\cos\xi$ by $\cos_{\rm
eff}\xi$, where the latter is a fitted function of the former.
Precisely speaking, by fitting the analytic calculation based upon
Eq.~(\ref{pi_survive_b}) \cite{Lipari:1993hd} to the Monte-Carlo
calculation, the relations between $\cos\xi$ and $\cos_{\rm eff}\xi$
can be found, which are tabulated in \cite{Gaisser:1997eu}.
Extrapolating such a relation, we find that $\cos_{\rm eff}\xi=0.05$
for $\cos\xi=0$. Using this $\cos_{\rm eff}\xi$ with
Eq.~(\ref{pi_survive_b}), we also calculate the atmospheric
$\nu_{\tau}$ flux. The result is compared with that obtained by the
full calculation (applying Eq.~(\ref{pi_survive_a})) as shown in
Fig.~\ref{comp}. Both results agree very well. Such an agreement
makes our calculation compelling and also validates the above
extrapolation on $\cos_{\rm eff}\xi$.

\begin{figure}
\includegraphics[width=7.in]{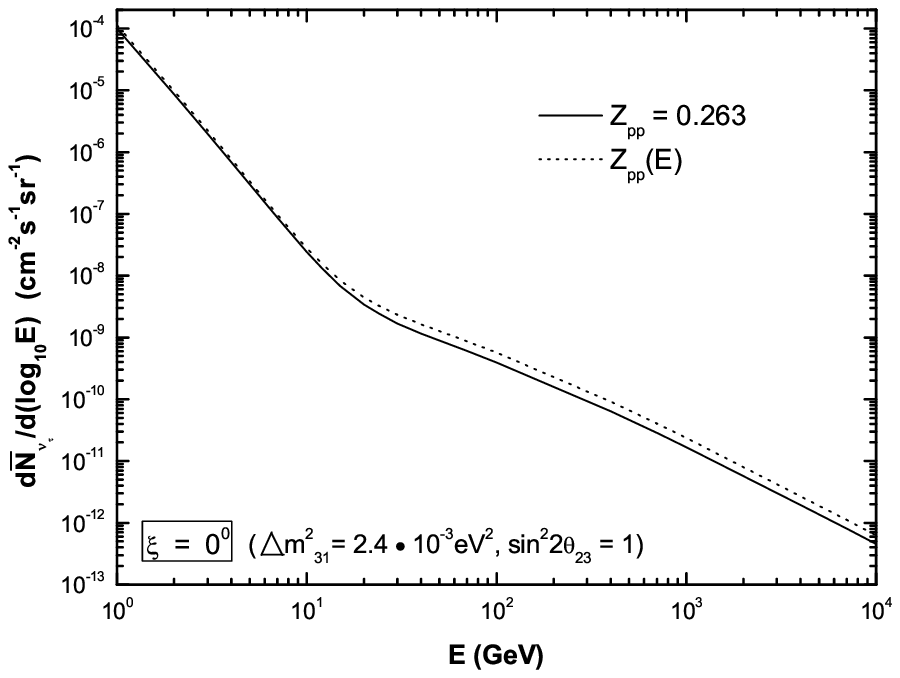}
\caption{The comparison of atmospheric $\nu_{\tau}$ fluxes
calculated from a constant $Z_{pp}$ \cite{COSMIC,Lipari:1993hd} and
an energy-dependent $Z_{pp}$ \cite{Gondolo:1995fq} for
$\xi=0^{\circ}$. } \label{z-depend1}
\end{figure}

\begin{figure}
\includegraphics[width=7.in]{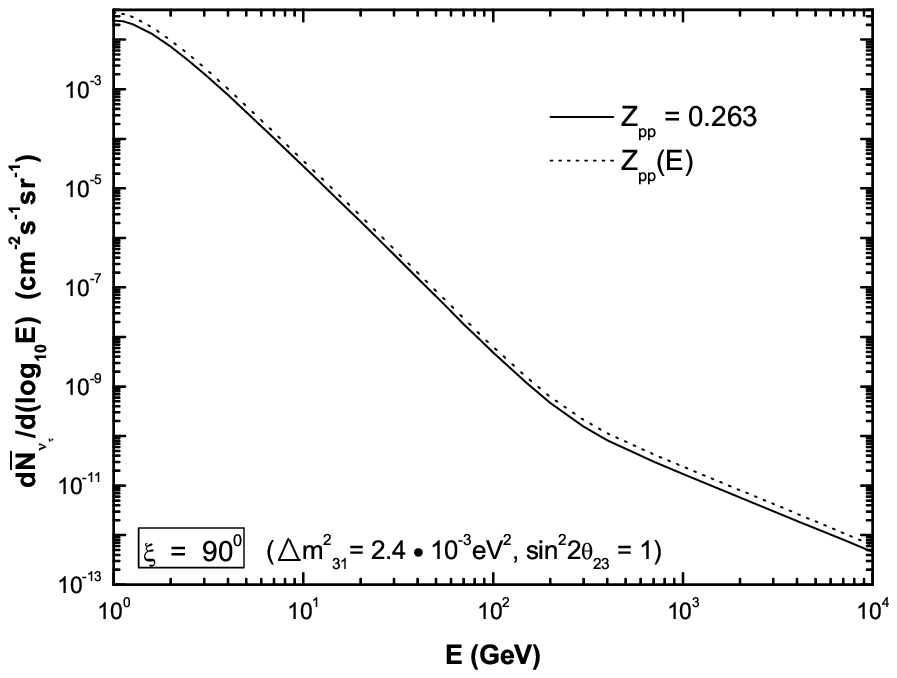}
\caption{The comparison of atmospheric $\nu_{\tau}$ fluxes
calculated from a constant $Z_{pp}$ \cite{COSMIC,Lipari:1993hd} and
an energy-dependent $Z_{pp}$ \cite{Gondolo:1995fq} for
$\xi=90^{\circ}$.} \label{z-depend2}
\end{figure}

We have so far computed the atmospheric neutrino flux with an energy
independent $Z$ moment, $Z_{pp}\equiv 1-\lambda_p/\Lambda_p=0.263$.
It is important to check the sensitivity of atmospheric $\nu_{\tau}$
flux on the values of $Z_{pp}$. We recall that different results for
$Z_{pp}$ are shown in Fig.~\ref{zpp}. At energies between $10^2$ GeV
and $10^3$ GeV, the values of $Z_{pp}$ generated by PYTHIA
\cite{pythia} slightly depend on the energy and roughly twice larger
than the value we have so far used for calculations. We check the
effect of $Z_{pp}$ by calculating the atmospheric $\nu_{\tau}$ flux
with the PYTHIA-generated $Z_{pp}$. The comparison of this result
with the earlier one obtained by setting $Z_{pp}=0.263$ is shown in
Fig.~\ref{z-depend1} for $\xi=0^{\circ}$ and Fig.~\ref{z-depend2}
for $\xi=90^{\circ}$. For $\xi=0^{\circ}$, two set of results do not
exhibit noticeable difference until $E\geq 10$ GeV. At $E=100$ GeV,
they differ by $45\%$. For $\xi=90^{\circ}$, two results differ by
$46\%$ at $E=1$ GeV while they differ by $29\%$ at $E=100$ GeV.
Obviously, the behavior of $Z_{pp}$ is one of the major
uncertainties for determining the atmospheric $\nu_{\tau}$ flux.
\subsubsection{ Muon-Decay contributions} We have stated that the
muon-decay contributions to $\nu_{\mu}$ is non-negligible for
neutrino energies less than $10$ GeV. Such $\nu_{\mu}$'s can
oscillate into $\nu_{\tau}$'s during their propagations in the
atmosphere. The calculation of such a flux according to
Eqs.~(\ref{atm-mu}) and (\ref{nu_mue}) is rather involved. However,
a simple approximation as presented below gives a rather accurate
result for this flux.

The calculation of $\nu_{\mu}$ spectrum due to muon decays requires
the knowledge of muon polarizations. The muon polarization however
depends on the ratio of muon momentum to the momentum of parent pion
or kaon as indicated by Eq.~(\ref{pol}). It is straightforward to
calculate the average muon polarization at any slant depth $X$
provided the energy spectrum of the parent pion or kaon is known at
that point. For the downward case ($\xi=0^{\circ}$), it is known
from the previous section that the muons are most likely produced at
around $14$ km from the ground detector. At that point, the pion and
kaon fluxes can be approximately parameterized as
$\phi_{\pi}(E_{\pi})=10^{-3.15}\cdot E_{\pi}^{-2.02}$ and
$\phi_{K}(E_K)=10^{-5.11}\cdot E_K^{-1.74}$ in units of
cm$^{-2}$s$^{-1}$sr$^{-1}$GeV$^{-1}$ for meson energies between $1$
and few tens of GeV. We do not distinguish $\pi^- (K^-)$ from
$\pi^+(K^+)$ in the above fittings. Although the spectra are charge
dependent, the resulting absolute values of $\mu^+$ polarization and
$\mu^-$ polarization differ by only $10\%$ for $E_{\mu}$ up to few
tens of GeV \cite{Lipari:1993hd}. From Eq.~(\ref{pol}), and the
above pion and kaon spectra, we obtain $\langle
P_{\mu^-}^{\pi}\rangle=0.35$, $\langle P_{\mu^-}^{K}\rangle=0.95$.
Therefore $\mu^-$ coming from the $\pi^-$ decays are $67\%$
right-handed polarized and $33\%$ left-handed polarized. On the
other hand, $\mu^-$ coming from $K^-$ decays are $98\%$ right-handed
polarized and only $2\%$ left-handed polarized. The muons produced
by meson decays lose energies before they decay into neutrinos. The
decay distribution for $\mu^-\to \nu_{\mu}$ is given by
Eq.~(\ref{distribution}). The average momentum fraction $\langle y
\rangle$ of muon neutrinos are $0.3$ and $0.4$ from decays of
right-handed $\mu^-$ and left-handed $\mu^-$ respectively. Following
a similar procedure, one can determine the polarization and decay
distributions of $\mu^+$. Finally, to calculate the spectrum of muon
neutrinos arising from muon decays, we use the approximation of
replacing $F_{\mu^{\pm}_s\to \nu_{\mu}}(E/E_{\mu})$ with
$\delta(E/E_{\mu}-\langle y \rangle)$ in Eq.~(\ref{nu_mue}).

To check the validity of the above approximation, we compare our
result on the fraction of muon decay contribution to the overall
$\nu_{\mu}$ flux with that given by Ref.~\cite{Lipari:1993hd} for
$\cos\xi=0.4$, i.e., $\xi=66^{\circ}$. At this zenith angle, most of
the muons are produced roughly $45$ km from the detector. The pion
and kaon fluxes at this point are fitted to be
$\phi_{\pi}(E_{\pi})=10^{-3.65}\cdot E_{\pi}^{-1.88}$ and
$\phi_{K}(E_K)=10^{-5.57}\cdot E_K^{-1.69}$ in units of
cm$^{-2}$s$^{-1}$sr$^{-1}$GeV$^{-1}$. This gives rise to $\langle
P_{\mu^-}^{\pi}\rangle=0.34$, $\langle P_{\mu^-}^{K}\rangle=0.94$.
Following the procedure in the downward case, we obtain the muon
neutrino flux from the muon decays. At $E=1$ GeV, the fraction of
muon-decay contributions to the overall $\nu_{\mu}$ flux is $44\%$
while the fraction decreases to $17\%$ at $E=10$ GeV. In
Ref.~\cite{Lipari:1993hd}, the corresponding fractions are $47\%$
and $18\%$ respectively. Both set of fractions agree rather well.

\begin{figure}
\includegraphics[width=7.in]{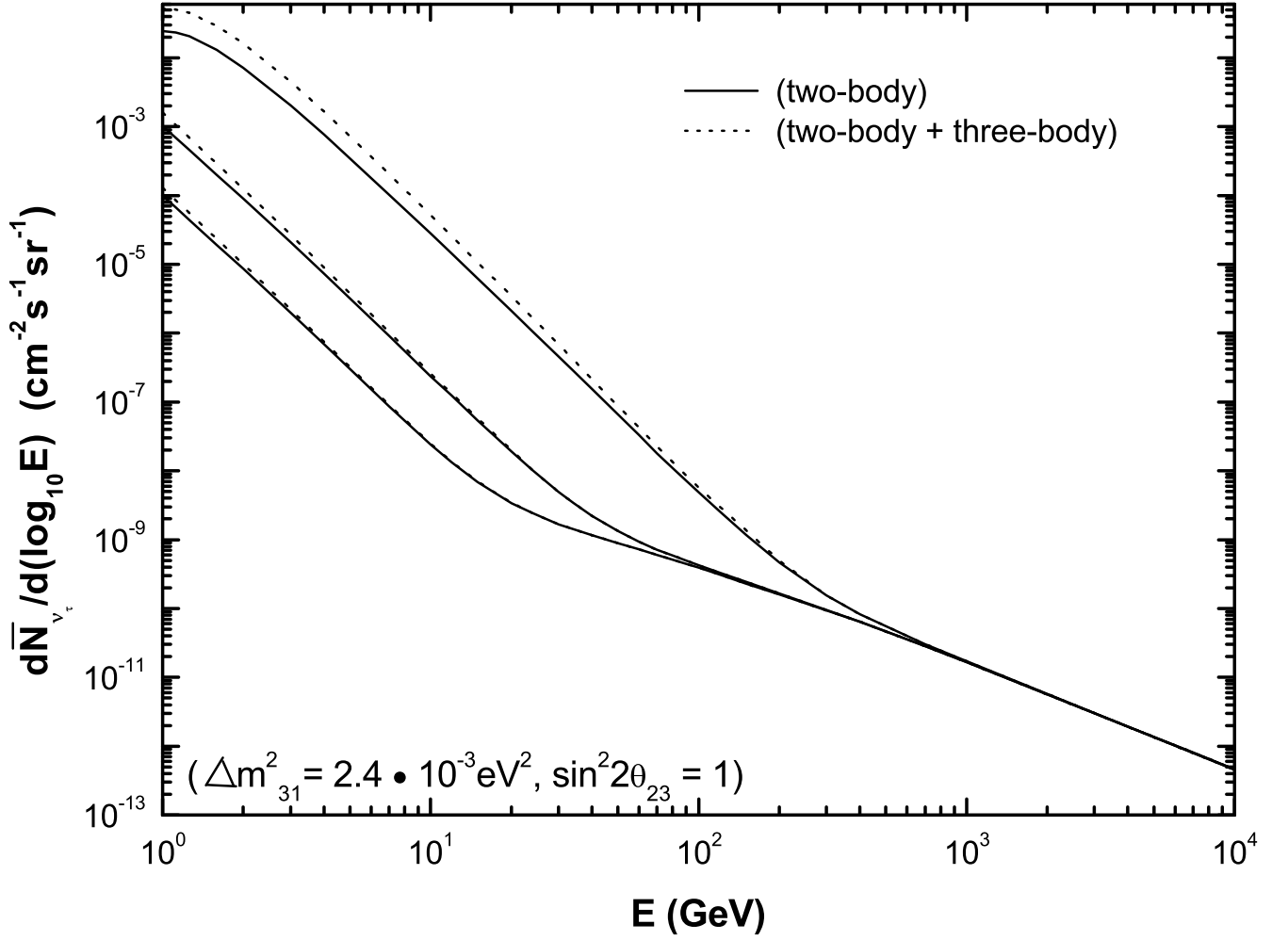}
\caption{The comparisons of atmospheric $\nu_{\tau}$ fluxes
resulting from the oscillations of $\nu_{\mu}$'s generated by
two-body and three-body decays with those resulting from the
oscillations of $\nu_{\mu}$'s generated only by two-body decays. The
comparisons are made for three zenith angles, $\cos\xi=0, \ 0.4$,
and $1$ (from top to bottom).} \label{two_three}
\end{figure}

Since our approximation works well for calculating the muon-decay
contributions to the atmospheric $\nu_{\mu}$ flux, we proceed to
calculate the resulting atmospheric $\nu_{\tau}$ flux with
Eq.~(\ref{oscillate}). Specifically we only need to include the
first term on the R.H.S. of Eq.~(\ref{oscillate}) because the second
term has already been included in the two-body decay contribution.
In Fig.~\ref{two_three}, those atmospheric $\nu_{\tau}$ fluxes
resulting from oscillations of $\nu_{\mu}$'s generated by both two-
and three-body decays (muon decays) are compared with those
resulting from the oscillations of $\nu_{\mu}$'s generated only by
two-body decays. As expected, the three-body decay contribution is
non-negligible for $E\leq 10$ GeV. Quantitatively, for
$\xi=0^{\circ}$ and $E= 1$ GeV, $24\%$ of the total atmospheric
$\nu_{\tau}$ flux is from the oscillations of $\nu_{\mu}$'s
originated from the muon decays. At $E= 10$ GeV, only $2.9\%$ of the
total atmospheric $\nu_{\tau}$ flux comes from the same source. For
$\xi=66^{\circ}$, the three-body decay contribution gives rise to
$36\%$ and $8.9\%$ of the total atmospheric $\nu_{\tau}$ flux at
$E=1$ GeV and $E=10$ GeV respectively. Finally, for
$\xi=90^{\circ}$, the three-body decay contribution to the total
atmospheric $\nu_{\tau}$ flux is most significant. It contributes to
$53\%$, $46\%$, and $39\%$ of the total atmospheric $\nu_{\tau}$
flux at $E=1$ GeV, $10$ GeV and $20$ GeV respectively. To calculate
the three-body decay contribution to $\nu_{\mu}$ flux at
$\xi=90^{\circ}$, we have used Eq.~(\ref{pi_survive_b}) for the
meson survival probability with $\cos_{\rm eff}\xi=0.05$ and a
overall factor $C\approx 1.40$ to fix the normalization of the flux
\cite{Gaisser:1997eu}.
\subsection{The Upward Atmospheric Tau Neutrino Flux} 
\begin{figure}
\includegraphics[width=7.in]{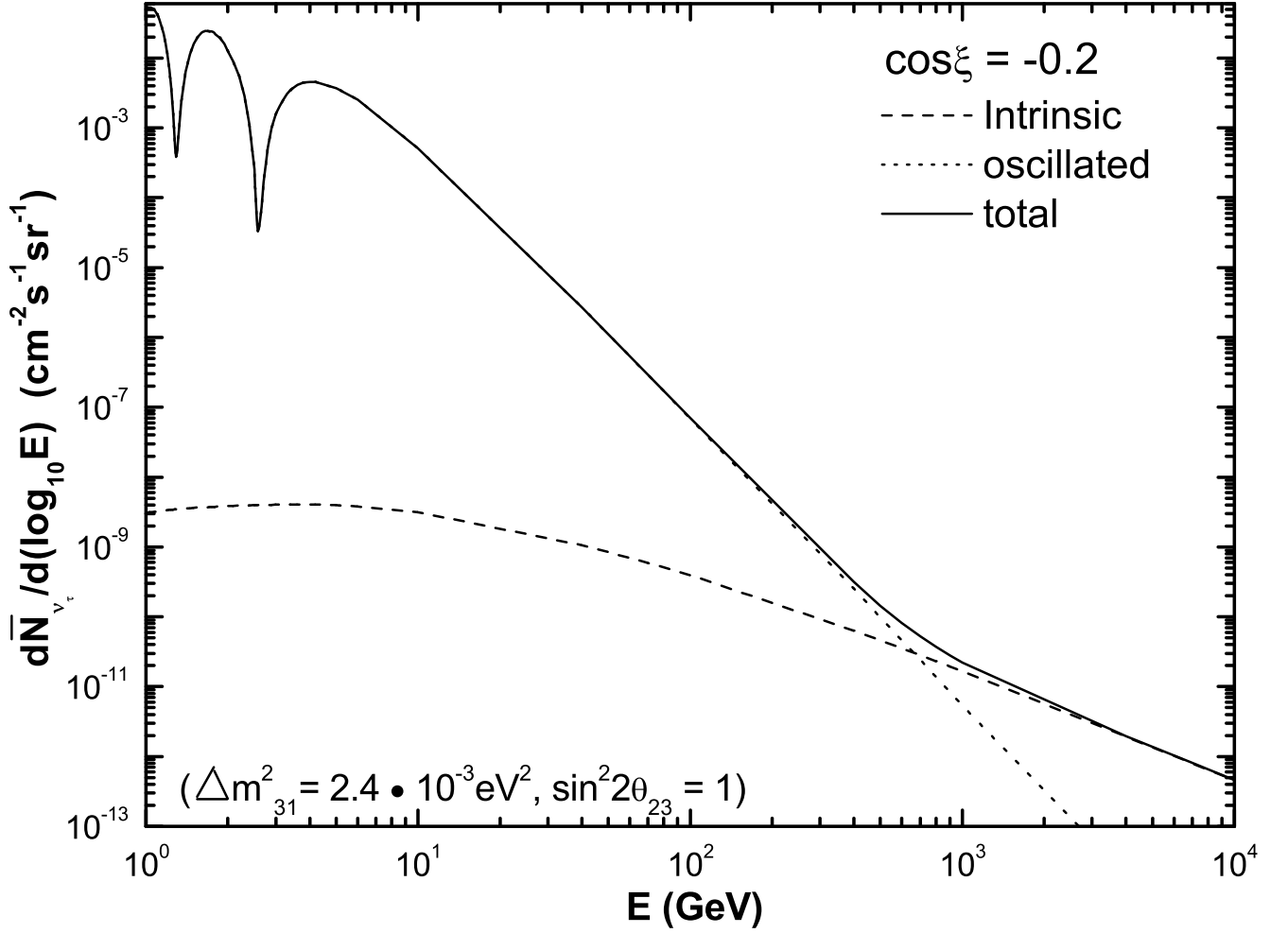}
\caption{The atmospheric $\nu_{\tau}$ flux (solid line) as a sum of
its oscillated (dotted line) and intrinsic (dashed line) components
for $\cos\xi=-0.2$ with $\sin^22\theta_{23}=1$, $\Delta
m_{31}^{2}=2.4\cdot 10^{-3}\, \, \, {\rm eV}^{2}$, and
$Z_{pp}=0.263$.  } \label{upward1}
\end{figure}

\begin{figure}
\includegraphics[width=7.in]{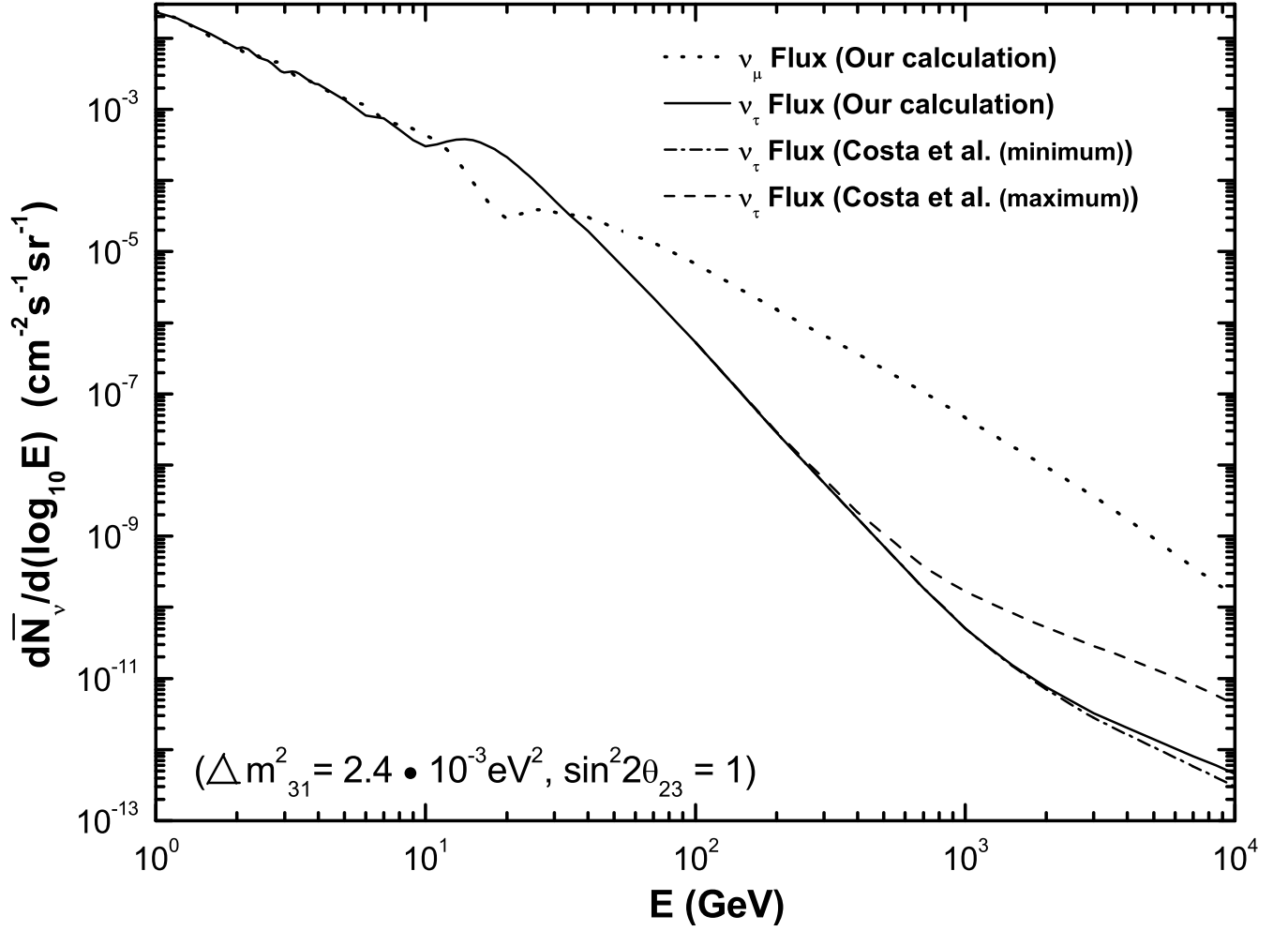}
\caption{The atmospheric $\nu_{\tau}$ flux averaged for $-1\leq
\cos\xi\leq -0.4$. The uncertainty of this flux due to the
uncertainty of intrinsic atmospheric $\nu_{\tau}$ flux is shown. The
corresponding $\nu_{\mu}$ flux is also plotted, which is equal to
the $\nu_{\tau}$ flux for $1\leq E/{\rm GeV}\leq 10$.}
\label{upward2}
\end{figure}

The upward
atmospheric $\nu_{\tau}$ fluxes are enhanced compared to those of
other directions since the average neutrino path lengths in such
case are larger. Therefore the observations of astrophysical tau
neutrinos in upward directions are subject to more serious
background problems. However, observing the atmospheric tau
neutrinos is interesting in its own right. The atmospheric tau
neutrino flux for $\cos\xi=-0.2$ is shown in Fig.~\ref{upward1}. The
effect of $\nu_{\mu}\to \nu_{\tau}$ oscillation is evident for below
TeV energies. This is seen from the crossing point of intrinsic and
oscillated atmospheric $\nu_{\tau}$ fluxes at $E\simeq 700$ GeV. The
atmospheric $\nu_{\tau}$ flux shows oscillatory behavior for $E\leq
10$ GeV. For $\cos\xi< -0.2$, such an oscillatory behavior is even
more significant. In such a case, it is more practical to study the
averaged flux. We average the atmospheric $\nu_{\tau}$ flux for the
zenith angle range $-1\leq \cos\xi \leq -0.4$, as shown in
Fig.~\ref{upward2}. Due to uncertainties of the intrinsic
atmospheric $\nu_{\tau}$ flux as discussed in
Ref.~\cite{Costa:2001fb}, the atmospheric $\nu_{\tau}$ flux taking
into account the oscillation effect also contains uncertainties
beginning at a few hundred GeV's. In the same figure, we also plot
the corresponding atmospheric $\nu_{\mu}$ flux. The $\nu_{\mu}$ and
$\nu_{\tau}$ fluxes are comparable for $E< 40$ GeV. In such a case,
the footprint of $\nu_{\tau}$ might be identified by studying the
energy spectra of shower events induced by neutrino interactions
\cite{Stanev:1999ki}. At $E=10^4$ GeV, the $\nu_{\mu}$ flux is
approximately $30$ times larger than the maximal $\nu_{\tau}$ flux.
We note that the maximal and minimal $\nu_{\tau}$ fluxes begin to
differ at $E=500$ GeV. At $E=1$ TeV, the maximal flux is $3$ times
larger than the minimal one. The ratio of maximal flux to the
minimal one increases to $14$ at $E=10$ TeV. We remark that the
upward atmospheric $\nu_{\tau}$ flux is also calculated in
Ref.~\cite{Stanev:1999ki} with $\sin^2 2\theta_{23}=1$ and $\Delta
m_{31}^2=10^{-2}, \ 10^{-2.5}, \ 10^{-3}$ eV$^2$ respectively. Here
we have done the calculation with the best fit value of $\sin^2
2\theta_{23}$ and $\Delta m_{31}^2$ taken from \cite{Ashie:2004mr}.
Furthermore we include the contribution of intrinsic atmospheric
$\nu_{\tau}$ flux and its associated uncertainties.
\section{Discussion and Conclusion}
\begin{figure}
\includegraphics[width=7.in]{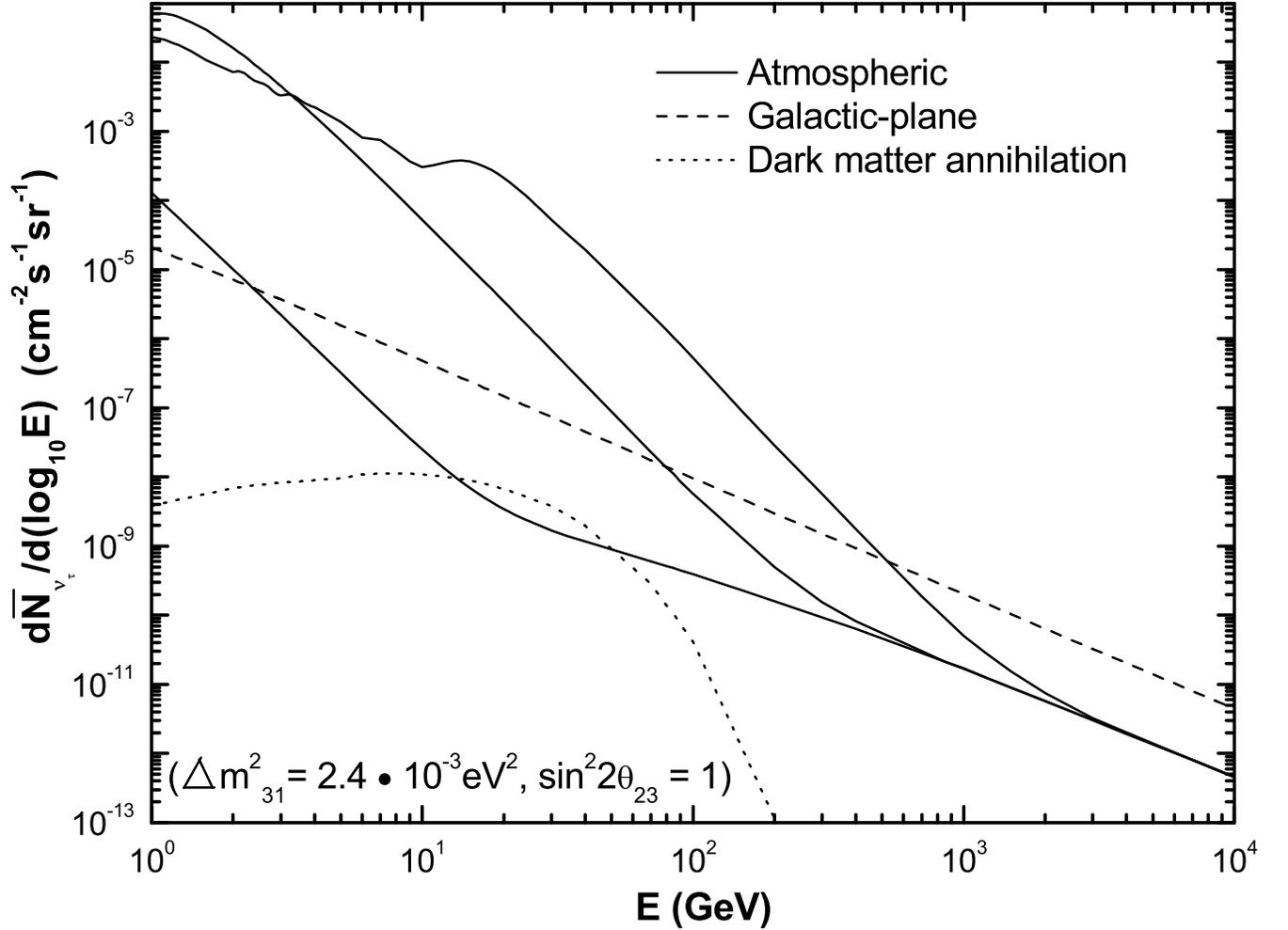}
\caption{The comparison of atmospheric $\nu_{\tau}$ fluxes with the
galactic-plane tau neutrino flux \cite{Athar:2004um} and the tau
neutrino flux due to the neutralino dark matter annihilations
\cite{Elsaesser:2004ck}. We have included downward ($\cos\xi=1$),
horizontal ($\cos\xi=0$) and upward ($-1\leq \cos\xi\leq -0.4$)
atmospheric $\nu_{\tau}$ fluxes for the comparison. At $E=1$ GeV,
the horizontal flux is the largest whereas the downward flux is the
smallest. } \label{various_tau}
\end{figure}

The understanding of atmospheric $\nu_{\tau}$ flux is important for
exploring the tau neutrino astronomy
\cite{Athar:2004pb,Athar:2004um}. As mentioned earlier, an
estimation of the atmospheric $\nu_{\tau}$ flux has been given in
Ref.~\cite{Athar:2004pb} while a detailed calculation of this flux
for zenith angles $0\leq \xi \leq 60^{\circ}$ is given in
Ref.~\cite{Athar:2004um}. In these works, comparisons of the
galactic-plane $\nu_{\tau}$ flux with the atmospheric $\nu_{\tau}$
flux are also made for illustrating the possibility of the tau
neutrino astronomy. Now that we have obtained a complete result of
the atmospheric $\nu_{\tau}$ flux for the entire zenith angle range,
we compare this flux with two astrophysical fluxes: the
galactic-plane tau neutrino flux just mentioned and the cosmological
$\nu_{\tau}$ flux due to neutralino dark matter annihilations
\cite{Elsaesser:2004ck}. The comparison is depicted in
Fig.~\ref{various_tau} where the flux of galactic-plane tau
neutrinos is taken from the calculation of Ref.~\cite{Athar:2004um}.
One can see that the galactic-plane $\nu_{\tau}$ flux dominates over
the downward ($\xi=0^{\circ}$) atmospheric $\nu_{\tau}$ flux for $E$
greater than a few GeV. Hence, in this direction, it is possible to
observe the flux of galactic-plane tau neutrinos in the GeV energy
range. For near horizontal directions, the atmospheric $\nu_{\tau}$
flux grows rapidly with zenith angles. Therefore, for
$\xi=90^{\circ}$, the energy threshold for galactic-plane tau
neutrino flux to dominate over its atmospheric counterpart is pushed
up to $E\simeq 100$ GeV. We further see that the galactic-plane
$\nu_{\tau}$ flux does not dominate the upward atmospheric
$\nu_{\tau}$ background ($-1\leq \cos\xi\leq -0.4$) until $E=500$
GeV. However, it is noteworthy that, in the muon neutrino case,
galactic-plane neutrino flux is overwhelmed by the atmospheric
background until $E_{\nu} > 10^6$ GeV \cite{Athar:2003nc}. Such a
difference between $\nu_{\mu}$ and $\nu_{\tau}$ shows the promise of
the tau neutrino astronomy in the GeV energy range as pointed out in
\cite{Athar:2004pb,Athar:2004um}. From Fig.~\ref{various_tau}, it is
also clear that the atmospheric $\nu_{\tau}$ flux is a
non-negligible background to the cosmological tau neutrino flux due
to neutralino dark matter annihilations \cite{Elsaesser:2004ck}. In
fact, two fluxes are comparable in the downward direction while the
atmospheric $\nu_{\tau}$ flux dominates in horizontal and upward
directions.

In summary, we have presented a semi-analytical calculation on the
atmospheric $\nu_{\tau}$ flux in the GeV to TeV energy range for
downward, upward, and horizontal directions. The atmospheric
$\nu_{\tau}$ flux at $\xi=90^{\circ}$ is two orders of magnitude
larger than the corresponding flux at $\xi=0^{\circ}$ for $1\leq
E/{\rm GeV}\leq 10$. On the other hand, the fluxes with zenith
angles between $0$ and $90$ degrees merge for $E\geq 700$ GeV,
provided that the intrinsic atmospheric $\nu_{\tau}$ flux is
calculated with perturbative QCD. Should one adopt a
non-perturbative model for the intrinsic $\nu_{\tau}$ flux, the
resulting $\nu_{\tau}$ fluxes on Earth at different zenith angles
would merge at an energy lower than $700$ GeV. We have observed that
the upward atmospheric $\nu_{\tau}$ fluxes show oscillatory
behaviors. For the averaged flux with $-1\leq \cos\xi\leq -0.4$, the
atmospheric $\nu_{\tau}$ flux is found to be comparable to the
atmospheric $\nu_{\mu}$ flux for $E< 40$ GeV. The comparison of this
flux with the horizontal atmospheric $\nu_{\tau}$ flux is also
interesting. Two fluxes are in fact comparable for $E< 10$ GeV. This
shows that the $\nu_{\mu}\to \nu_{\tau}$ oscillation is already
quite significant in the horizontal direction for such an energy
range. Nevertheless, the upward atmospheric $\nu_{\tau}$ flux takes
over from $E\geq 10$ GeV until $E\simeq 3$ TeV where two fluxes
merge again. Concerning the uncertainties in our calculations, we
have studied the dependencies of atmospheric $\nu_{\tau}$ flux on
the $Z$ moment $Z_{pp}$ for representative zenith angles
$\xi=0^{\circ}$ and $\xi=90^{\circ}$. We have also discussed in
detail the uncertainty of intrinsic atmospheric $\nu_{\tau}$ flux
due to different models for charm hadron productions. The
consequence of such a uncertainty on the determination of oscillated
$\nu_{\tau}$ flux is studied as well. Concerning the technique for
calculating the atmospheric $\nu_{\tau}$ flux from large zenith
angles, we have verified the validity of using $\cos_{\rm eff}\xi$
in Eq.~(\ref{pi_survive_b}) to calculate the atmospheric
$\nu_{\tau}$ flux for $\xi
> 60^{\circ}$.
In particular, we have extrapolated the results in
Ref.~\cite{Gaisser:1997eu} to $\xi=90^{\circ}$ and demonstrate that
the choice $\cos_{\rm eff}(\xi=90^{\circ})=0.05$ reproduces well the
atmospheric $\nu_{\tau}$ flux obtained by a full calculation using
Eq.~(\ref{pi_survive_a}).
\section*{Acknowledgements}
This work is supported by the National Science Council of Taiwan
under the grant number NSC 93-2112-M-009-001.
\pagebreak
\pagebreak
\pagebreak
\pagebreak
\pagebreak
\pagebreak
\pagebreak
\pagebreak
\pagebreak
\pagebreak
\pagebreak
\pagebreak
\pagebreak
\end{document}